\documentclass[12pt]{article}
\usepackage{epsf}
\usepackage[utf8]{inputenc}
\usepackage[english]{babel}

\begin{document}

\title{
{\bf Effective transverse radius of nucleon in high-energy elastic diffractive scattering}}
\author{A.A. Godizov\thanks{E-mail: anton.godizov@gmail.com}\\
{\small {\it Institute for High Energy Physics, NRC ``Kurchatov Institute'', 142281 Protvino, Russia}}}
\date{}
\maketitle

\begin{abstract}
High-energy elastic diffraction of nucleons is considered in the framework of the simplest Regge-eikonal approximation. It is demonstrated explicitly that the effective 
transverse radius of nucleon in this nonperturbative regime is $\sim 0.2\div 0.3$ fm and much less than the transverse size of the diffractive interaction region. 
\end{abstract}

\section*{Introduction}

Elastic diffractive scattering of hadrons is one of the most interesting and important areas of high-energy hadron physics: in $pp$ collisions the fraction of elastic 
diffraction events in the total number of events is very high (from more than 15\% at the ISR to about 25\% at the LHC). However, the main problem related to this sector of 
strong interaction physics is the fact that the characteristic distances for diffractive interaction of hadrons are of order 1 fm, so that perturbative QCD is inapplicable. 
Hence, one has to search for some approaches not related to pQCD directly, to provide at least qualitative description of the corresponding high-energy observables. 

One of the most natural theoretical frameworks which helps to deal with the nonperturbative sector of hadron physics is Regge theory wherein interaction of hadrons is 
described in terms of exchanges by reggeons (off-mass-shell and off-spin-shell composite particles). In paper \cite{godizov} a simple Regge-eikonal model for high-energy 
elastic diffraction of nucleons was examined. In this model, the standard eikonal representation of the nonflip scattering amplitude \cite{collins}, 
\begin{equation}
\label{eikrepr}
T_{el}(s,t) = 4\pi s\int_0^{\infty}db^2J_0(b\sqrt{-t})\frac{e^{2i\delta(s,b)}-1}{2i}\,,
\end{equation}
$$
\delta(s,b) = \frac{1}{16\pi s}\int_0^{\infty}d(-t)J_0(b\sqrt{-t})\delta(s,t)
$$
(here $s$ and $t$ are the Mandelstam variables and $b$ is the impact parameter), is exploited together with the single-reggeon-exchange approximation to the eikonal (Born 
amplitude) in the kinematic range $s\gg \{m_p^2,|t|\}$: 
\begin{equation}
\label{eikphen}
\delta(s,t) = \delta_{\rm P}(s,t) = \left(i+{\rm tg}\frac{\pi(\alpha_{\rm P}(t)-1)}{2}\right){\Gamma_{\rm P}}^2(t)\left(\frac{s}{s_0}\right)^{\alpha_{\rm P}(t)},
\end{equation}
where $s_0 = 1$ GeV$^2$ and $\alpha_{\rm P}(t)$ is the Regge trajectory of pomeron (a $C$-even reggeon which absolutely dominates over other reggeons at the SPS, Tevatron, 
and LHC energies). More detailes, regarding the Regge-eikonal approach, can be found in \cite{collins} or, partly, in the Appendix.

In \cite{godizov} the unknown functions $\alpha_{\rm P}(t)$ and $\Gamma_{\rm P}(t)$ are considered independent and treated with the help of some simple test 
parametrizations which are fitted to the data. Formally, the used parametrizations have provided a satisfactory description of the available experimental data at 
$\sqrt{s}>$ 500 GeV and $-t<$ 2 GeV$^2$. However, a question 
emerges about possible correlation between the behavior of $\alpha_{\rm P}(t)$ and $\Gamma_{\rm P}(t)$ at low $t$, since both the functions have a strong impact on the 
$t$-behavior of the scattering amplitude. 

Such a correlation exists and can be taken into account explicitly. As a consequence, it becomes possible to extract from the data a valuable information concerning the 
effective transverse radius of nucleon in the nonperturbative regime of high-energy diffractive scattering.

\section*{Structure of the pomeron Regge residue}

First of all, we should note that in the framework of the Regge-eikonal approach the pomeron exchange contribution into the eikonal of nucleon-nucleon elastic scattering 
appears as 
\begin{equation}
\label{eikphen2}
\delta_{\rm P}(s,t) = \left(i+{\rm tg}\frac{\pi(\alpha_{\rm P}(t)-1)}{2}\right)g^2_{\rm P}(t)\;\pi\alpha'_{\rm P}(t)\left(\frac{s}{2s_0}\right)^{\alpha_{\rm P}(t)},
\end{equation}
where $\alpha'_{\rm P}(t)$ originates from the pomeron propagator (see the Appendix) and, in general, the factor $2^{-\alpha_{\rm P}(t)}\pi\alpha'_{\rm P}(t)$ is not 
related to the pomeron-nucleon coupling. In literature this factor is usually included into the corresponding Regge residue \cite{collins}: 
$g^2_{\rm P}(t)2^{-\alpha_{\rm P}(t)}\pi\alpha'_{\rm P}(t)\equiv {\Gamma_{\rm P}}^2(t)$.

If to consider $g_{\rm P}(t)$ as a nontrivial unknown function of $t$, then expressions (\ref{eikphen}) and (\ref{eikphen2}) are equivalent from the standpoint of 
description of data. However, the replacement $g_{\rm P}(t)\to \Gamma_{\rm P}(t)$ degrades the physical transparency of the model since the shortened form (\ref{eikphen}) 
ignores the evident correlation between the behavior of the pomeron Regge residue and the pomeron Regge trajectory. Moreover, if the $t$-dependence of $g_{\rm P}(t)$ is 
weak at low $t$, then usage of (\ref{eikphen}) instead of (\ref{eikphen2}) leads to loss of physical information. 

Treating the high-energy elastic scattering of nucleons in the same way as the lepton-proton elastic scattering, we define the quantity $g_{\rm P}(0)$ as the effective 
``pomeron charge'' of nucleon, while the ratio $g_{\rm P}(t)/g_{\rm P}(0)$ should be considered as the nucleon ``pomeron form factor''. By analogy with the 
extraction of the proton charge radius from the proton charge form factor, one can extract the effective transverse ``pomeron radius'' of nucleon from $g_{\rm P}(t)$. 
Hence, possible weak $t$-dependence of this quantity at low $t$ could be interpreted as the effective transverse (quasi-)pointlikeness of nucleon in the high-energy 
diffractive scattering regime. 

As well, let us remind that although QCD itself does not predict the behavior of $\alpha_{\rm P}(t)$ in the diffraction domain ($0<-t<2$ GeV$^2$), this function 
is expected to satisfy the following conditions \cite{collins,kearney}:
\begin{equation}
\label{gluon}
\frac{d^n\alpha_{\rm P}}{dt^n}>0\;\;(n=1,2,...\,;\;\;t<0)\,,\;\;\;\lim_{t\to -\infty}\alpha_{\rm P}(t) = 1\,.
\end{equation}
The first condition originates from the dispersion relations for Regge trajectories (if not more than one subtraction is needed), and the second one follows from the 
natural presumption that at high values of the transferred momentum the exchange by pomeron turns, due to asymptotic freedom, into the exchange by 2 noninteracting gluons 
which can be considered in the same way as the exchange by 2 photons. At high energies such Born amplitudes behave as $\sim s^1$ \cite{wu}.

Thus, if the effective transverse radius of nucleon is small, then, in view of restrictions (\ref{gluon}), the $t$-dependence of eikonal (\ref{eikphen2}) is determined mainly 
by the Herglotz function $\alpha_{\rm P}(t)$. Consequently, the quality of description of the differential cross-section $\frac{d\sigma}{dt} = \frac{|T(s,t)|^2}{16\pi s^2}$ 
becomes extremely sensitive to the quantitative behavior of $\alpha_{\rm P}(t)$ at low negative $t$. Such a sensitivity implies that the procedure of fitting 
$\alpha_{\rm P}(t)$ and $g_{\rm P}(0)$ to experimental angular distributions in a wide enough kinematic range could be considered as implicit extraction of these 
quantities from the data.

\section*{Fitting to the experimental data}

Let us reconsider the model from \cite{godizov}, having singled out the factor $2^{-\alpha_{\rm P}(t)}\pi\alpha'_{\rm P}(t)$ in the Regge residue, as in (\ref{eikphen2}):
\begin{equation}
\label{pomeron}
\alpha_{\rm P}(t) = 1+\frac{\alpha_{\rm P}(0)-1}{1-\frac{t}{\tau_a}}\,,\;\;\;\;\Gamma_{\rm P}(t)\to g_{\rm P}(t)=\frac{g_{\rm P}(0)}{(1-a_gt)^2}\,,
\end{equation}

\begin{figure}[ht]
\vskip -0.5cm
\epsfxsize=8.2cm\epsfysize=8.2cm\epsffile{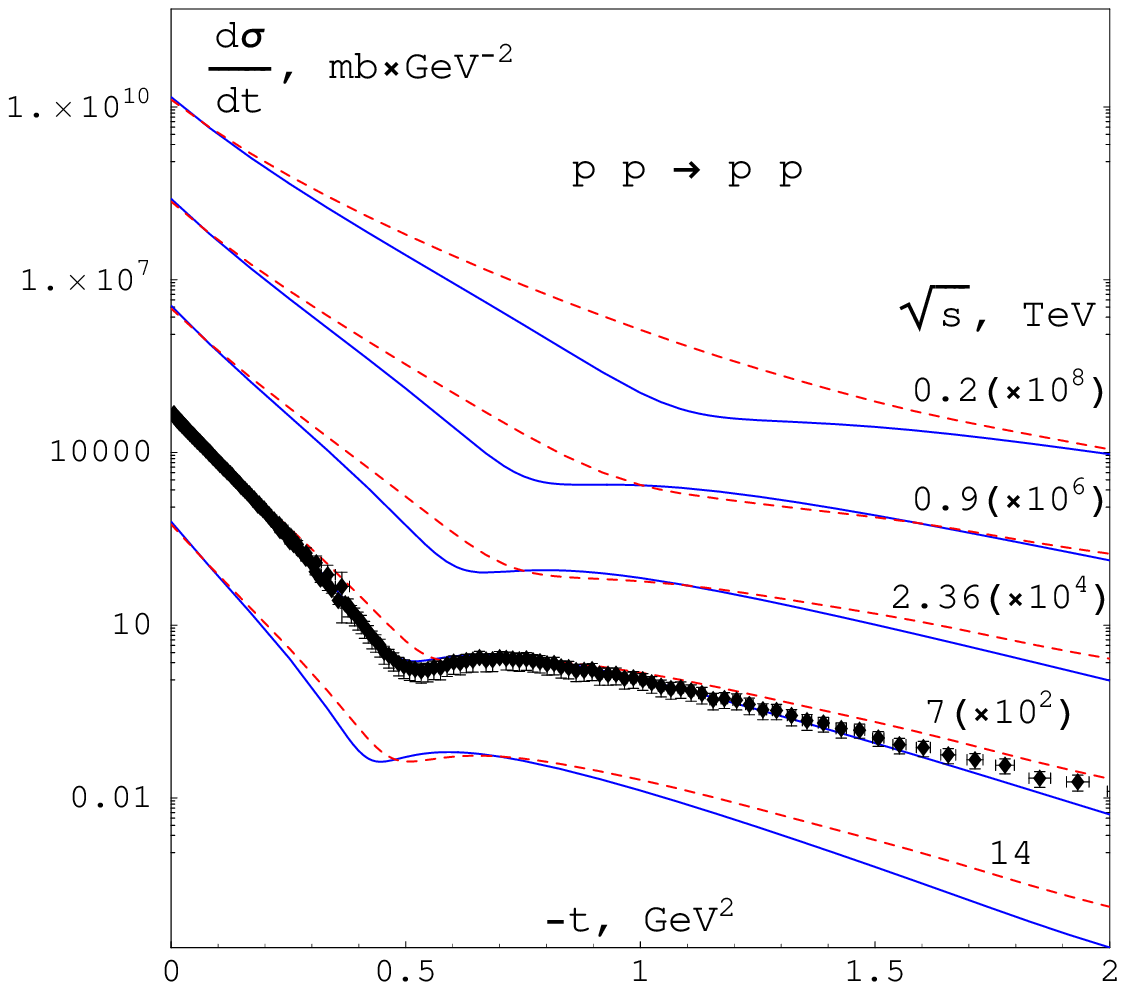}
\vskip -8.25cm
\hskip 8.5cm
\epsfxsize=8.2cm\epsfysize=8.2cm\epsffile{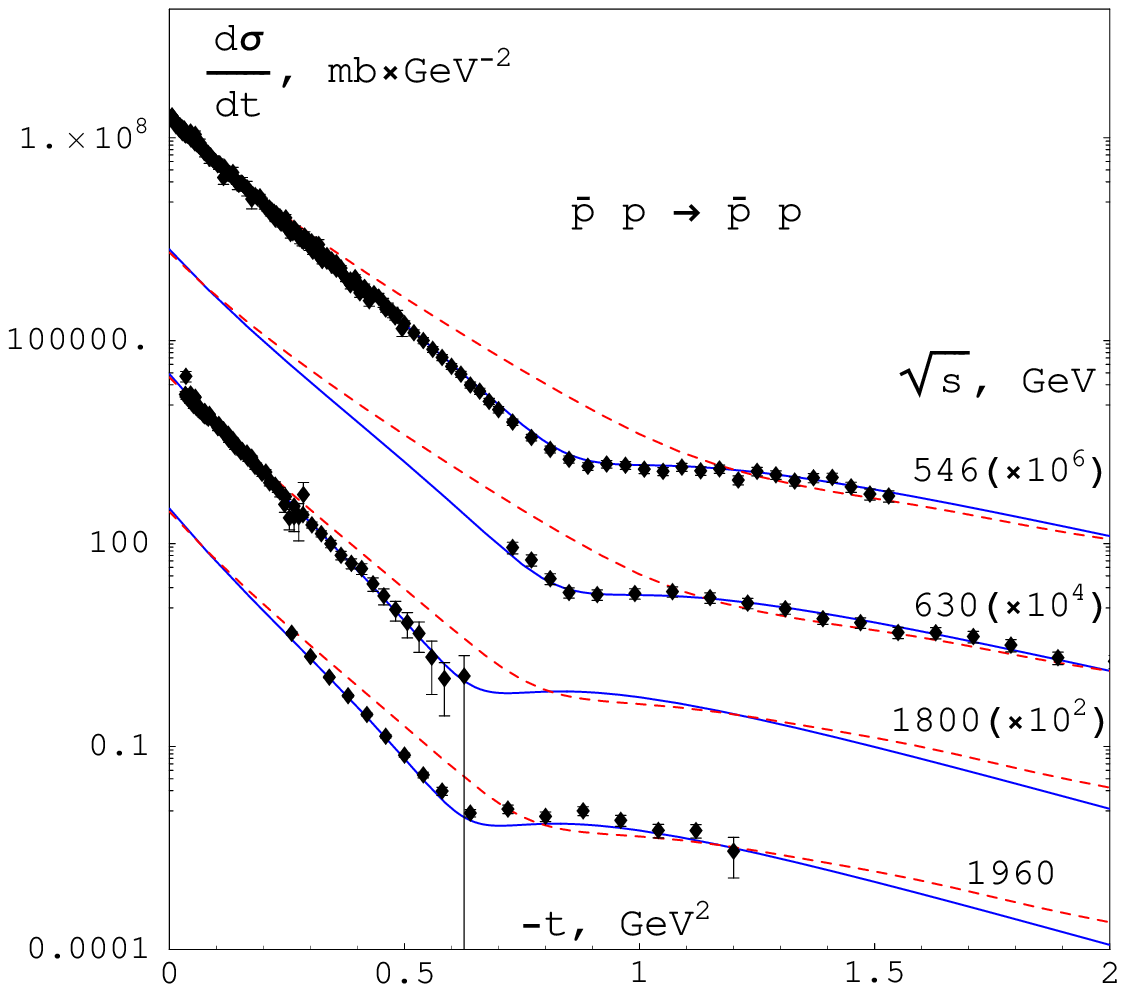}
\vskip -0.2cm
\epsfxsize=8.2cm\epsfysize=8.2cm\epsffile{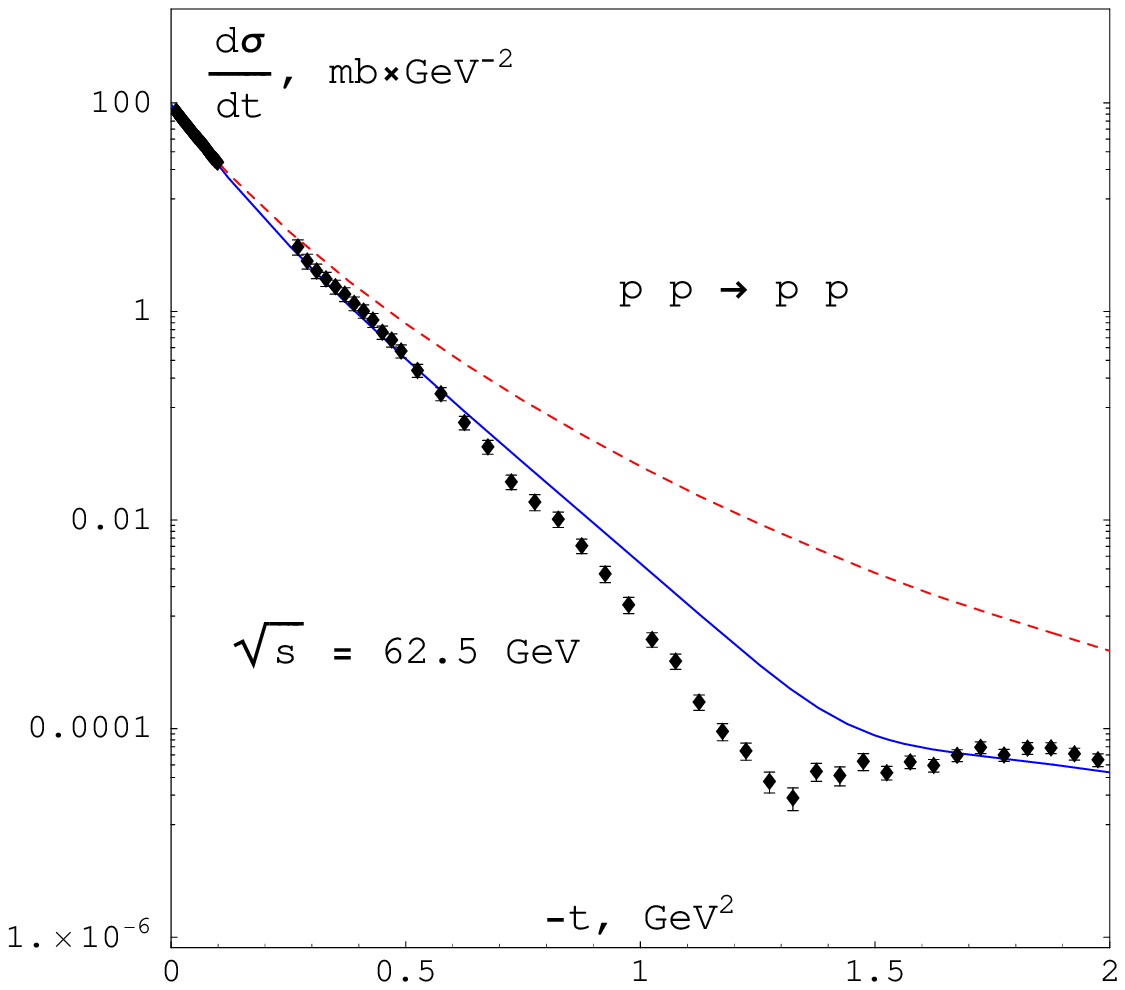}
\vskip -8.1cm
\hskip 9.1cm
\epsfxsize=8.0cm\epsfysize=8.0cm\epsffile{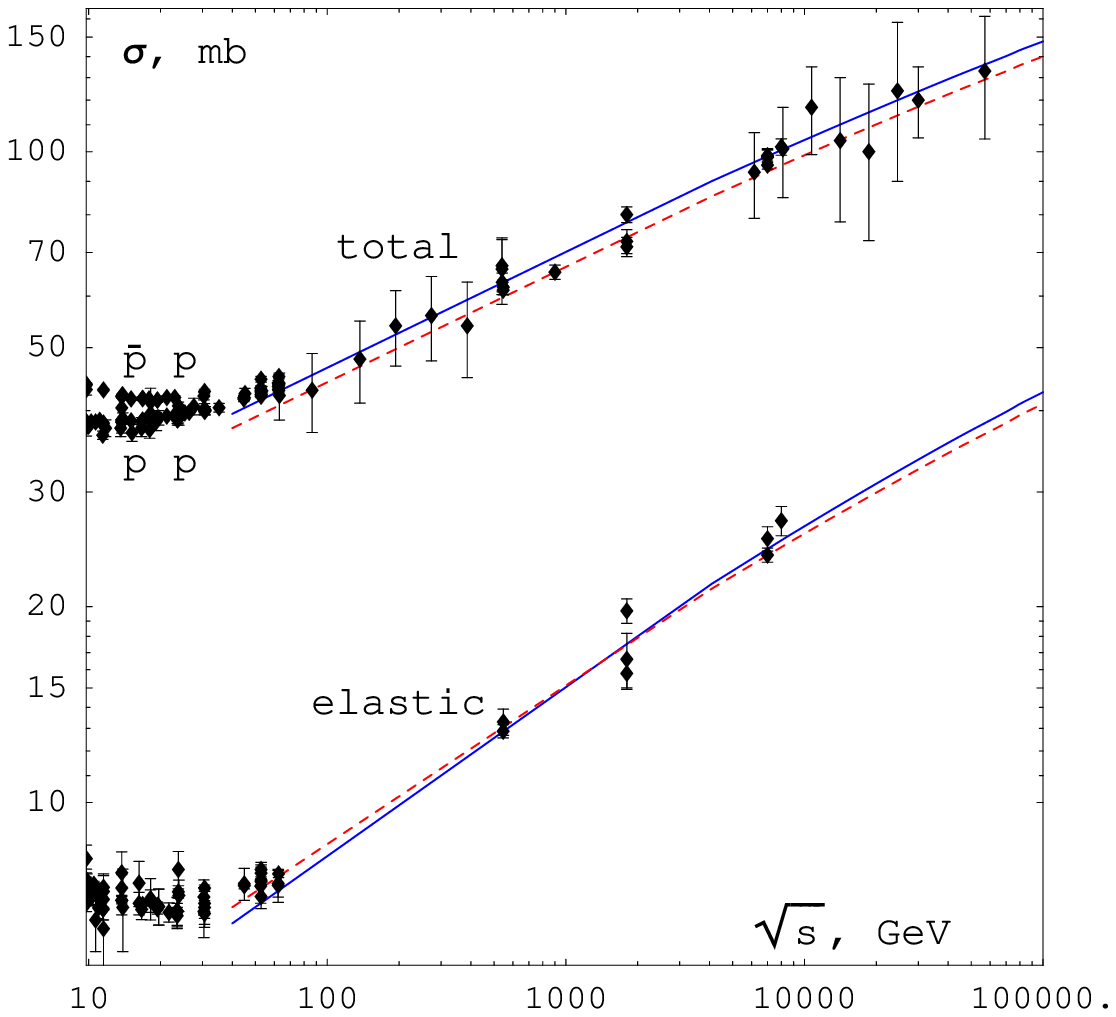}
\vskip -0.5cm
\caption{The high-energy evolution of the nucleon-nucleon elastic scattering observables. The dashed lines correspond to the value $a_g=0$ instead of $a_g=0.23$ GeV$^{-2}$.}
\label{diff}
\end{figure}

The results of fitting $\alpha_{\rm P}(t)$ and $g_{\rm P}(t)$ to the experimental differential cross-sections at $\sqrt{s}>$ 500 GeV and 0.005 GeV$^2<-t<$ 2 GeV$^2$ 
\cite{diffexp} are presented in Tabs. \ref{tab1}, \ref{tab2}, and \ref{tab3} and Fig. \ref{diff}. The deviation of the model predictions from the $pp$ elastic scattering 
data in the dip region at $\sqrt{s}=$ 62.5 GeV \cite{isr62} can be explained by the noticeable contribution of secondary reggeons into the real part of the eikonal. 
Detailed discussion of this matter can be found in \cite{godizov}.
\begin{table}[ht]
\begin{center}
\begin{tabular}{|l|l|}
\hline
\bf Parameter          & \bf Value                   \\
\hline
$\alpha_{\rm P}(0)-1$  & $0.109\pm 0.017$            \\
$\tau_a$               & $(0.535\pm 0.057)$ GeV$^2$    \\
$g_{\rm P}(0)$         & $(13.8\pm 2.3)$ GeV          \\
$a_g$                  & $(0.23\pm 0.07)$ GeV$^{-2}$ \\
\hline
\end{tabular}
\end{center}
\vskip -0.2cm
\caption{The parameter values obtained via fitting to the high-energy differential cross-section data.}
\label{tab1}
\end{table}
\vskip -0.5cm
\begin{table}[ht]
\begin{center}
\begin{tabular}{|l|l|l|}
\hline
$\sqrt{s}$, GeV                        & \bf Number of points &  $\chi^2$   \\
\hline
 546 ($\bar p\,p$; UA1, UA4, CDF)      & 231                  &  253        \\
 630 ($\bar p\,p$; UA4)                & 17                   &   11        \\
1800 ($\bar p\,p$; E710)               & 51                   &   16        \\
7000 ($p\,p$; TOTEM, ATLAS)            & 201                  &  188        \\
\hline
\bf Total                              & 500                  &  468        \\
\hline
\end{tabular}
\end{center}
\vskip -0.2cm
\caption{The quality of description of the data \cite{diffexp} on the angular distributions of nucleon-nucleon scattering.}
\label{tab2}
\end{table}

The D0 data \cite{d0diff} were not included into the fitting procedure, since they have a normalization uncertainty about 14.4\%. If to multiply them by factor 0.92, the 
description quality becomes much better (see Tab. \ref{tab4}). The same can be said regarding the unrenormalized CDF data at $\sqrt{s}=$ 1800 GeV \cite{cdf} which are 
inconsistent with the E-710 data.
\begin{table}[ht]
\begin{center}
\begin{tabular}{|l|l|l|l|}
\hline
$\sqrt{s}$, GeV & $\sigma_{tot}$, mb & $\sigma_{el}$, mb & $B$, GeV$^{-2}$ \\
\hline
    62.5        & $ 43.0\pm  4.4$    & $ 7.4\pm 1.2$     & $ 14.5\pm 0.8$  \\
   200          & $ 53.3\pm  3.8$    & $10.2\pm 1.2$     & $ 16.0\pm 0.8$  \\
   546          & $ 63.8\pm  3.3$    & $13.2\pm 1.0$     & $ 17.3\pm 0.9$  \\
  1800          & $ 78.5\pm  3.4$    & $17.8\pm 1.0$     & $ 19.1\pm 1.0$  \\
  7000          & $ 98.4\pm  5.4$    & $24.5\pm 1.8$     & $ 21.4\pm 1.1$  \\
  8000          & $100.5\pm  5.7$    & $25.3\pm 1.9$     & $ 21.6\pm 1.1$  \\
 13000          & $108.6\pm  6.9$    & $28.2\pm 2.5$     & $ 22.5\pm 1.2$  \\
 14000          & $109.9\pm  7.1$    & $28.6\pm 2.5$     & $ 22.6\pm 1.2$  \\
 32000          & $124.9\pm  9.7$    & $34.1\pm 3.7$     & $ 24.2\pm 1.4$  \\
100000          & $148.0\pm 14.1$    & $42.8\pm 5.6$     & $ 26.6\pm 1.7$  \\
\hline
\end{tabular}
\end{center}
\vskip -0.2cm
\caption{Predictions for the $pp$ total and elastic cross-sections and the forward logarithmic slope of the corresponding differential cross-sections.}
\label{tab3}
\end{table}

\begin{table}[ht]
\begin{center}
\begin{tabular}{|l|l|l|}
\hline
$\sqrt{s}$, GeV                             & \bf Number of points &  $\chi^2$   \\
\hline
1960 ($\bar p\,p$; D0)                      & 17                   &   55        \\
1960 ($\bar p\,p$; D0, multiplied by 0.92)  & 17                   &   29        \\
1800 ($\bar p\,p$; CDF)                     & 26                   &  178        \\
1800 ($\bar p\,p$; CDF, multiplied by 0.88) & 26                   &   45        \\
\hline
\end{tabular}
\end{center}
\vskip -0.2cm
\caption{The quality of description of the data \cite{d0diff,cdf} not included into the fitting procedure.}
\label{tab4}
\end{table}

\section*{Justification for usage of the parametrization for $\alpha_{\rm P}(t)$}

Before making any physical conclusions on the fitting results, we should discuss our choice of parametrization for $\alpha_{\rm P}(t)$. 

First, the essential nonlinearity of $\alpha_{\rm P}(t)$ from (\ref{pomeron}), which, in fact, is not related to this concrete expression, but follows from restrictions 
(\ref{gluon}), seems to be in contradiction with the observed approximate linearity of the Chew-Frautschi plots for secondary reggeons. 
However, we would like to point out that such a linear behavior of secondaries takes place at $t>0$ and, 
in principle, it is not guaranteed for $t<0$. Moreover, it was determined in the framework of the BFKL approach \cite{kwiecinski} that secondary Regge trajectories behave as 
$\alpha_{\rm R}(t)=\left(\frac{8\alpha_s(\sqrt{-t})}{3\pi}\right)^{1/2}+O(\alpha_s^{5/6}(\sqrt{-t}))$ at high negative $t$, where $\alpha_s(\mu)$ is the QCD running 
coupling. Hence, in view of rather high slopes of the corresponding Chew-Frautschi plots in the resonance region, we have a very simple alternative: either secondary Regge 
trajectories are essentially nonlinear in the diffraction domain, or they are not monotonic functions in the interval $-\infty<t<0$. Detailed discussion of this matter can be 
found in \cite{petrov}.

Second, we should ascertain that the obtained smallness of $a_g$ (see Tab. \ref{tab1}) is not related directly to the chosen specific parametrization for $\alpha_{\rm P}(t)$, 
since this is crucial for the main conclusion of the paper. If to consider a simple generalization of (\ref{pomeron}), 
\begin{equation}
\label{pomeron2}
\alpha^{(k)}_{\rm P}(t) = 1+\frac{\alpha^{(k)}_{\rm P}(0)-1}{\left(1-\frac{t}{\tau^{(k)}_a}\right)^k}\,,\;\;\;\;
g^{(k)}_{\rm P}(t)=\frac{g^{(k)}_{\rm P}(0)}{(1-a^{(k)}_gt)^2}\;,
\end{equation}
where $k$ takes on integer and half-integer values (parametrization (\ref{pomeron}) corresponds to $k=1$), then it is possible to provide a satisfactory description of the 
data in a few cases of this series (see Tab. \ref{tab5}). The description for $k=1/2$ and $k\ge 3$ is unsatisfactory. For any $k>1$, we find that $a^{(k)}_g<a^{(1)}_g$. 
Thus, the conclusion we make below is not related to specific form (\ref{pomeron}) of $\alpha_{\rm P}(t)$ only.

At the very end of this section we would like to note that although in this paper we restricted ourselves by the simplest test parametrizations, the quantities 
$\alpha_{\rm P}(t)$ and $g_{\rm P}(t)$ should, in general, be treated as unknown functions in the framework of the considered model. Namely, any expression for 
$\alpha_{\rm P}(t)$ could be used which is analytic at $t<0$ and satisfies conditions (\ref{gluon}). Certainly, it should provide a satisfactory description of the available 
data as well.

\begin{table}[ht]
\begin{center}
\begin{tabular}{|l|l|l|l|}
\hline
\bf Parameter                & $k = 3/2$                     & $k = 2$                        & $k = 5/2$                     \\
\hline
$\alpha^{(k)}_{\rm P}(0)-1$  & $0.108\pm 0.017$              & $0.108\pm 0.017$               & $0.108\pm 0.017$              \\
$\tau^{(k)}_a$               & $(0.708\pm 0.079)$ GeV$^2$    & $(0.874\pm 0.100)$ GeV$^2$     & $(1.040\pm 0.125)$ GeV$^2$    \\
$g^{(k)}_{\rm P}(0)$         & $(13.1\pm 2.2)$ GeV           & $(12.6\pm 2.0)$ GeV            & $(12.3\pm 2.0)$ GeV           \\
$a^{(k)}_g$                  & $(0.167\pm 0.075)$ GeV$^{-2}$ & $(0.120\pm 0.077)$ GeV$^{-2}$  & $(0.085\pm 0.080)$ GeV$^{-2}$ \\
\hline
$\chi^2/D.o.F.$              & 0.915                         & 0.93                           & 0.97                          \\
\hline
\end{tabular}
\end{center}
\vskip -0.2cm
\caption{The fitting results for different variants of parametrization (\ref{pomeron2}).}
\label{tab5}
\end{table}

\section*{Discussion}

Now let us analyze the produced outcomes. 

So, the description of the nucleon-nucleon diffractive pattern in the considered kinematic range is satisfactory. The replacement of the fitted value of $a_g$ by $a_g=0$, 
which implies neglection of the nucleon shape, disfigures the differential cross-sections (see the dashed lines in Fig. \ref{diff}). However, this distortion decreases 
with energy and becomes not catastrophic already at the LHC energies. It is a consequence of the fact that although we did not fix the nucleon transverse radius, it has 
turned out rather small. Indeed, the form factor $g_{\rm P}(t)/g_{\rm P}(0)=(1-a_gt)^{-2}$ corresponds to a certain effective transverse distribution in the impact parameter 
representation: $f(b)=(4\pi a_g^3)^{-1}bK_1(b/a_g)$, where $K_1(x)$ is the modified Bessel function. The effective transverse radius of nucleon obtained via average over 
this distribution, $\sqrt{<b^2>}\sim 0.2\div 0.3$ fm, is noticeably smaller than the effective transverse size of the diffractive interaction region (in the 
considered interval of the collision energy $\sqrt{2B}>1$ fm, $B$ is the forward logarithmic slope of $d\sigma/dt$ which increases with energy). Such a difference could be 
interpreted as if pomeron was coupled to a very small zone inside nucleon. Hence, we come to the main conclusion:
\begin{itemize}
\item The quantitative evolution of the nucleon-nucleon elastic diffractive scattering observables at ultrahigh energies is determined mainly by the behavior of the pomeron 
Regge trajectory $\alpha_{\rm P}(t)$ and the value of the effective ``pomeron charge'' $g_{\rm P}(0)$ of nucleon and rather weakly depends on the nucleon shape.
\end{itemize}

The examined single-reggeon-exchange eikonal approximation is expected to be valid at least in the interval 0.2 TeV $\le\sqrt{s}\le 14$ TeV. Therefore, the forthcoming 
TOTEM measurements (as well as desirable analogous measurements at the RHIC) could decrease the uncertainties of the model degrees of freedom and, thus, improve the 
phenomenological estimations as for the nucleon effective transverse radius, so for the pomeron Regge trajectory and the pomeron charge of nucleon.
\subsection*{Acknowledgments} The author thanks V.A. Petrov and R.A. Ryutin for discussion.

\section*{Appendix. Regge approximation for Born amplitude}

The eikonal representation (\ref{eikrepr}) itself implies just a replacement of the unknown function of two variables, $T(s,t)$, by another one, $\delta(s,t)$. The key 
assumption is that the eikonal is proportional (with high accuracy) to some effective relativistic quasi-potential of two-hadron interaction. According to the Van Hove 
interpretation of such a quasi-potential as the ``sum'' over all single-meson exchanges in the $t$-channel \cite{vanhove}, the eikonal can be represented as 
$$
\delta = \sum_{j=0}^{\infty}\sum_{M_j}J^{(f_1,j,M_j)}_{\alpha_1...\alpha_j}(p_1,\Delta)\frac{D_{(M_j)}^{\alpha_1...\alpha_j,\beta_1...\beta_j}(\Delta)}
{m_j^2-\Delta^2}J^{(f_2,j,M_j)}_{\beta_1...\beta_j}(p_2,-\Delta)\,,
\eqno{(\rm A.1)}
$$
where $\frac{D_{(M_j)}^{\alpha_1...\alpha_j,\beta_1...\beta_j}}{m_j^2-\Delta^2}$ is the propagator of spin-$j$ meson particle, $m_j^2=M_j^2-i M_j\Gamma_j$ ($M_j$ and 
$\Gamma_j$ are the meson mass and decay width), $J^{(f,j,M_j)}_{\alpha_1...\alpha_j}$ are the corresponding meson currents of interacting hadrons (index $f$ denotes the sort 
of the hadron), $\Delta$ is the transferred 4-momentum, $p_1$ and $p_2$ are the 4-momenta of the incoming particles, and symbol $\sum_{M_j}$ denotes the summing over all of 
spin-$j$ mesons with different masses (which, in what further, will be transformed into the summing over reggeons). 

Obviously, in the kinematic range $(p_1+p_2)^2\equiv s\gg \{|p_{1,2}^2|,|\Delta^2|,|(p_{1,2}\Delta)|\}$ the eikonal can be approximated by the following expression: 
$$
\delta(s,t) = \sum_{j=0}^{\infty}\sum_{M_j}\frac{h^{(j,M_j)}(t)}{m_j^2-t}\left(\frac{s}{2}\right)^j,
\eqno{(\rm A.2)}
$$
where $t\equiv\Delta^2$, $h^{(j,M_j)}(t)\equiv f^{(f_1,j,M_j)}(p_1^2,\Delta^2,(p_1\Delta))f^{(f_2,j,M_j)}(p_2^2,\Delta^2,-(p_2\Delta))$, and $f^{(f,j,M_j)}$ is 
the structure function at the tensor structure $p_{\alpha_1}...p_{\alpha_j}$ in the current $J^{(f,j,M_j)}_{\alpha_1...\alpha_j}(p,\Delta)$. 

Now let us introduce the single-meson-exchange amplitudes of definite signature:
$$
\delta_\pm(s,t) = \sum_{j=0}^{\infty}\sum_{M_j}(1\pm e^{-i\pi j})\;\frac{h^{(j,M_j)}(t)}{m_j^2-t}\left(\frac{s}{2}\right)^j.
\eqno{(\rm A.3)}
$$
If $m_j^2$ and $h^{(j,M_j)}(t)$ at even and odd $j$ are the values of some analytic functions which are holomorphic at ${\rm Re}\,j >-\frac{1}{2}$ and behave as 
$O(e^{k|j|})$, $k<\pi$, at $j\to\infty$, then, under the Carlson theorem \cite{carlson}, the unilocal analytic continuation of (\rm A.3) into the region of complex $j$ is 
possible (the Regge hypothesis \cite{collins}). We denote these functions by $m_\pm^2(j)$ and $h_\pm(j,t)$, respectively. Via the Sommerfeld-Watson transform 
\cite{collins,zommer}, we replace the sum over $j$ in (\rm A.3) by the integral over the contour $C$ encircling the real positive half-axis on the complex $j$-plane, 
including the point $j = 0$, in such a way that the half-axis is on the right:
$$
\delta_\pm(s,t) = \frac{1}{2i}\oint_C\frac{dj}{\sin(\pi j)}\sum_{m_\pm}(\mp1-e^{-i\pi j})
\frac{h_\pm(j,t)}{m_\pm^{2}(j)-t}\left(\frac{s}{2}\right)^j.
\eqno{(\rm A.4)}
$$
According to our assumption, the only sources of the integrand singularities in the region ${\rm Re}\,j >-\frac{1}{2}$ are the zeros of the functions $\sin(\pi j)$ and 
$m_\pm^2(j)-t$. Hence, deforming the contour $C$ to the axis ${\rm Re}\,j = -\frac{1}{2}$ (the behavior of $h_\pm(j,t)$ at $j\to\infty$ 
and ${\rm Re}\,j >-\frac{1}{2}$ should allow such a deformation), we obtain 
$$
\delta_\pm(s,t) = \frac{1}{2i}\int_{-\frac{1}{2}-i\infty}^{-\frac{1}{2}+i\infty}\frac{dj}{\sin(\pi j)}\sum_{m_\pm}(\mp1-e^{-i\pi j})
\frac{h_\pm(j,t)}{m_\pm^{2}(j)-t}\left(\frac{s}{2}\right)^j \; + 
$$
$$
+ \; \sum_n\frac{\mp1-e^{-i\pi \alpha_n^\pm(t)}}{\sin(\pi\alpha_n^\pm(t))}\frac{\pi d\alpha_n^\pm(t)}{dt}\,
h_\pm(\alpha_n^\pm(t),t)\left(\frac{s}{2}\right)^{\alpha_n^\pm(t)},
\eqno{(\rm A.5)}
$$
where the functions $\alpha_n^\pm(t)$ are the roots of the equations $m_\pm^2(j)\;-\;t = 0$ and, thus, they correspond to the eikonal poles in the complex 
$j$-plane. These poles are called Regge poles, and the functions $\alpha_n^\pm(t)$ are called Regge trajectories. 

At high enough values of $s$ the background integral contribution is negligible. As the functions $h_\pm$ can be factorized into two factors related to each of the 
interacting hadrons, so we come to the following expression for the eikonal:
$$
\delta_\pm(s,t) = \sum_n\xi_\pm(\alpha_n^\pm(t))\;g_n^{(1)\pm}(t)g_n^{(2)\pm}(t)\;\frac{\pi d\alpha_n^\pm}{dt}\left(\frac{s}{2s_0}\right)^{\alpha_n^\pm(t)},
\eqno{(\rm A.6)}
$$
where $s_0$ is some scale determined {\it a priori} (for example, $s_0 = 1$ GeV$^2$) and related directly to the factors $g_n^{(i)}(t)$ which should be interpreted as the 
effective couplings of reggeons to the colliding particles. 
$\xi_\pm(\alpha)$ are the so-called reggeon signature factors: $\xi_+(\alpha)=i+{\rm tg}\frac{\pi(\alpha-1)}{2}$ and $\xi_-(\alpha)=i-{\rm ctg}\frac{\pi(\alpha-1)}{2}$.

The last formula (which is valid, as well, for inelastic scattering $2\to 2$ and for reactions with off-shell particles), together with the eikonal representation 
(\ref{eikrepr}) of the scattering amplitude, is the essence of the Regge-eikonal approach \cite{collins}.

\end{document}